\documentclass{PoS}
 \usepackage[usenames,dvipsnames]{pstricks}
 \usepackage{epsfig}
 \usepackage{pst-grad}
 \usepackage{pst-plot}
\title{Resummation of clustering logarithms for non-global QCD observables}
\ShortTitle{Resummation of CLs for NG QCD observables}
\author{\speaker{Yazid DELENDA}          \\
        Universit\'{e} Hadj Lakhdar - Batna, Algeria\\
        E-mail: \email{yazid.delenda@gmail.com}}
\author{Kamel KHELIFA-KERFA\\
        Universit\'{e} des Sciences et de la Technologie Houari Boumediene - Algiers, Algeria\\
        E-mail: \email{kamel.kkhelifa@gmail.com}}

\newcommand{\di}{\mathrm{d}}

\abstract{We address the problem of resumming leading clustering logs in QCD jet observables defined using the $k_t$, CA and SISCone algorithms. We specifically choose the jet mass distribution as an example and calculate up to $\mathcal{O}(\alpha_s^4)$ clustering-log terms in the series expansion at single-log accuracy. These terms are found to exhibit a pattern of exponentiation and  we are thus able to perform an all-orders analytical resummation for the clustering logs. We also numerically calculate the non-global logs at large $N_c$. We show that our result for the resummation of clustering logs is a very good analytical approximation to the numerical result obtained using a specialised Monte Carlo program.}

\FullConference{XXI International Workshop on Deep-Inelastic Scattering and Related Subjects\\
                22-26 April, 2013\\
                Marseille, France}

\begin{document}

\section{Introduction}

With the large beam energy of the LHC massive high-$p_t$ ``background'' QCD jets are formed that resemble the structure of ``signal'' jets resulting from the collimated decay of heavy boosted particles. Thus  the field of jet substructure \cite{substructure} has recently become very active with the aim of providing clean signal/backgound discrimination of such boosted heavy jets. In this regard jet shapes have, along other jet substructure techniques, played an essential role in improving the power of the already available discrimination methods as well as offering new ones \cite{substructure2}.

Additionally jet shapes form an indispensable tool in testing and tuning many Monte Carlo (MC) event generators \cite{Sjostrand:2006za}. The tuning process is often amenable to large uncertainties that are dominated by non-perturbative effects inherent in QCD. It is vital that such tuning be as accurate as possible since any mis-appointed effects may spoil future comparisons to measurements which heavily rely on the said MCs. For instance, perturbative calculations typically involve large uncertainties due to neglected observable-dependent sub-leading effects which could then potentially be labelled as due to universal non-perturbative effects. The latter are then used in various other comparisons to measurements, e.g. extraction of properties of heavy particles.

Progress in analytical calculations, alongside MC estimates, could provide a clean extraction and discrimination of both perturbative and non-perturbative aspects and hence contribute towards eliminating the problem of mis-tuning. For instance, knowing the dependence of jet-shape distributions on jet algorithms and jet parameters, such as jet radii $R$, had led to the development of the concept of optimal jet algorithms and jet radius \cite{Dassalmag}. Furthermore analytical calculations often have a definite control of the size and impact of neglected sub-leading effects, thus giving more confidence on the clean extraction of various components of measured cross-sections.

Jet/event shape distributions have received substantial progress in the last few years on the analytical side. For instance, the resummation of the thrust distribution is now available up to NNNLL accuracy in the exponent of the distribution \cite{N3LL}. Furthermore matching the resummed analytical distribution to fixed-order results is achieved up to NNNLO \cite{N3LO}. Progress has also been made in disentangling various non-perturbative components such as the underlying event (UE) and the hadronisation by means of analytical computations of, e.g. power corrections to the jet-$p_t$ \cite{Dassalmag}.

The success of the resummation of the thrust distribution cannot unfortunately be extended to all jet/event shape distributions, the most obvious obstacle being the non-global nature of many jet shapes . For instance the jet mass distribution of a high-$p_t$ jet, which we hereafter pick as an explicit example, has large non-global logs (NGLs) which are currently only resummable numerically in the large-$N_c$ limit \footnote{The authors of a recent paper \cite{ngsol} claim to have calculated these numerically in full $N_c$ .} at NLL level \cite{ngl1}. This means that even a full NLL resummation is not currently available which could lead to large uncertainties due to neglected subleading $\mathcal{O}(1/N_c)$ terms.

Another complication that non-global jet shapes have is the impact of jet algorithms on their distributions. The effect of jet clustering ($k_t$ algorithm) on the resummation of NGLs was first studied in ref. \cite{Appleby:2002ke} for the  gaps-between-jets energy ($E_t$) flow distribution. It was later shown in ref. \cite{BD05} that another class of single logs, which we refer to as clustering logs (CLs), also show up for non-global jet quantities. These logs were first analytically calculated at fixed order and numerically resummed to all orders in ref. \cite{BD05}. An analytical approximation to the all-orders resummation of CLs was presented in ref. \cite{Delenda:2006nf}. Moreover, in the latter reference it was shown that the impact of NGLs is far more reduced than suggested in ref. \cite{Appleby:2002ke} for the case of inter-jet $E_t$ flow.

In ref. \cite{Delenda:2006nf} the CLs were resummed in the exponent of the distribution as a power series in the jet radius $R$ which rapidly converges, particularly for small $R$. It was thus sufficient to only calculate a couple of terms in the series to generate an accurate distribution which compared well with the numerical result. We note that in this case (inter-jet $E_t$ flow) collinear emissions to the jets do not change the gap energy, and thus the leading logs are single logs. In the case of the jet mass observable, however, collinear emissions to the jet contribute to its mass, giving rise to large double logs. Furthermore, as we shall show, the CLs are not resummed into a power series in $R$.

Here we aim at analytically resumming the CLs for the jet mass distribution, which were first computed in ref. \cite{BDKM} at two-loop, to all orders at single log accuracy in various jet algorithms. We only explicitly present the results for the $k_t$ algorithm, as it is the only algorithm currently implemented in the numerical MC program of \cite{ngl1, Appleby:2002ke, Delenda:2006nf}. For completeness we also provide the numerically resummed NGLs factor in the large-$N_c$ limit for the $k_t$ and anti-$k_t$ algorithms. We show how different jet algorithms affect different parts of the distribution (global, non-global, and clustering). In the next section we discuss the role of jet clustering by showing how both CLs and NGLs get generated at two-gluon (primary and secondary) emission level. We then write down the all-orders resummed distribution in various algorithms. We compare our analytical resummation of CLs (in the $k_t$ algorithm) with the output of the MC program of refs. \cite{ngl1, Appleby:2002ke, Delenda:2006nf} and also show the numerical result of NGLs. This leads us to a discussion about the concept of optimal jet algorithms and jet radius from which we draw our conclusions.

\section{Non-global and clustering logs at leading order}

\subsection{Jet mass and jet algorithms}

The normalised invariant jet mass is defined by:
\begin{equation}
\rho = \left(\sum_{j\in \textrm{jet}} p_j\right)^2/\left(\sum_i E_i\right)^2 ,
\end{equation}
where the sum in the nominator extends over all particles in the measured jet defined using a suitable jet algorithm, and the sum in the denominator runs over all particles in the event. Our aim is the resummation of the normalised single inclusive integrated jet mass distribution:
\begin{equation}
\Sigma(R^2/\rho) = \int_0^{\rho} \frac{1}{\sigma} \frac{\di\sigma}{\di\rho'} \di\rho' ,
\end{equation}
where $R$ is the jet radius. For simplicity, and without loss of generality, we consider dijet production in $e^+e^-$ annihilation where we measure the mass of one of the jets leaving the other jet unmeasured.

We consider jets defined using four famous jet algorithms, one of cone-type and three of sequential-recombination-type. For the latter each pair of particles ($ij$) in the event is assigned a distance measure $d_{ij} = \min(k_{ti}^{p},k_{tj}^{p}) \theta_{ij}^2$, with $\theta_{ij}^2=\delta \eta_{ij}^2+\delta \phi_{ij}^2$, that depends on the transverse momenta $k_{ti}$, rapidities $\eta_i$ and azimuths $\phi_i$ of the particles. Additionally each individual particle ($i$) has a beam-distance measure $d_i=k_{ti}^p\,R^2$. The values of $p$, being $-1$, $0$ and $1$, represent the three well-known algorithms, anti-$k_t$, CA and $k_t$ respectively \cite{antikt,CA,kt}. The algorithm sequentially recombines the closest (in terms of the said distances) particles first by adding their four-momenta. If a particle (pseudo-jet) is closest to the beam then it is considered a jet and is removed from the list of initial pseudo-jets. For the SISCone algorithm \cite{SISCONE}, on the other hand, the clustering proceeds in two steps. Firstly, the algorithm searches for \emph{all} stable protojets in a seedless way, and secondly a split-merge procedure is applied on overlapped protojets.

\subsection{Non-global and clustering logs at two-loop}

\begin{figure}[ht]
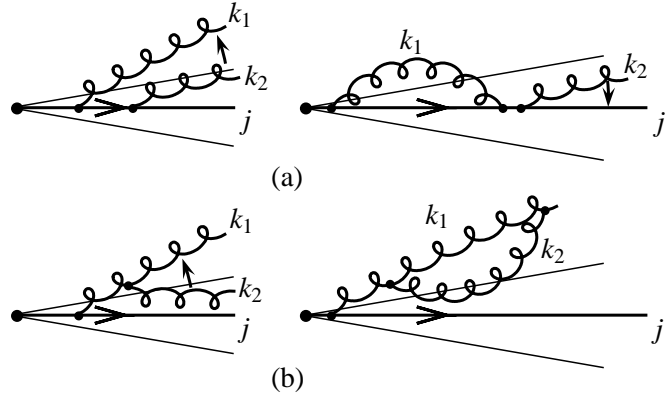

\begin{center}\scalebox{1}{
}\end{center}
\caption{\label{fig:Yazid_Delenda_feyn} Feynman diagrams contributing to (a) CLs and (b) NGLs.}
\end{figure}

In this section we illustrate how CLs emerge at two-loop for $k_t$ clustering. We also discuss how, compared to anti-$k_t$ clustering, NGLs are significantly reduced in the former clustering. We first note that in order to extract the single logs it is sufficient to consider the emission of two energy-ordered soft gluons $k_1$ and $k_2$ off the hard quark $j$ initiating the measured jet, with $k_{t2}\ll k_{t1}\ll Q$ and $Q$ the hard scale. Second we recall that, for primary emissions, particle configurations that give rise to double logs in the anti-$k_t$ case are not altered by the $k_t$ clustering. New configurations resulting from the latter clustering and giving rise to single CLs are depicted in Fig. \ref{fig:Yazid_Delenda_feyn}(a). In this case the relevant $k_t$-algorithm distances satisfy $\theta_{12}<\theta_{2j}<R<\theta_{1j}$ such that particle $k_2$ is clustered with $k_1$ when the latter is real, while it is clustered with the hard jet $j$ when $k_1$ is virtual. A real-virtual miscancellation at the cross-section level then leads to the appearance of single logs of the form $C_F^2\alpha_s^2 L^2$, with $L=\ln(R^2/\rho)$, in the jet mass distribution.

We note that in the case of the anti-$k_t$ algorithm the particle $k_2$ in the same diagram is always recombined with the hard jet $j$, leading to a complete real-virtual cancellation and hence the absence of CLs in this case.

Next we consider secondary emissions, depicted in Fig. \ref{fig:Yazid_Delenda_feyn}(b), where particle $k_2$ is emitted off the harder particle $k_1$. If $k_2$ is recombined with $k_1$ by the jet algorithm, then the real and virtual contributions, shown in Fig. \ref{fig:Yazid_Delenda_feyn}(b), cancel out completely giving no NGLs. However if $k_2$ is recombined with the hard jet $j$, then a real-virtual miscancellation occurs and NGLs of the form $C_F C_A \alpha_s^2 L^2$ are generated. For the anti-$k_t$ algorithm the said miscancellation takes place provided that $\theta_{2j}<R<\theta_{1j}$, while for the $k_t$ algorithm we require, in addition to $\theta_{2j}<R<\theta_{1j}$, that $\theta_{2j} < \theta_{12}$. The extra condition of the $k_t$ algorithm ($\theta_{2j} < \theta_{12}$) restricts the available phase-space for the contribution of NGLs since it requires the two particles, $k_1$ and $k_2$, to be at wide angles from each other. Recalling that NGLs are at maximum effect when $k_1$ and $k_2$ are collinear, this means that, compared to the anti-$k_t$ algorithm, the impact of NGLs is reduced in the $k_t$ algorithm.

In summary, while for the anti-$k_t$ algorithm there are no CLs in the distribution and NGLs are at maximum impact,  for the $k_t$ algorithm both CLs and NGLs are present, with the latter being largely reduced in effect. In the next section we shall illustrate this by plotting various components of the distribution for the two algorithms.

\section{Jet mass distribution at all orders}

The NLL resummed jet mass distribution can be expressed in the following form:
\begin{equation}\label{eq:YDres}
\Sigma^{\mathrm{algo}} = \Sigma_{\mathrm{glob}} \times S_{\mathrm{NG}}^{\mathrm{algo}} \times C_{\mathrm{clus}}^{\mathrm{algo}} ,
\end{equation}
where $\Sigma_{\mathrm{glob}} = \exp\left(Lg_1(\alpha_sL)+g_2(\alpha_sL)\right)$ is the famous Sudakov form factor that is common between global and non-global observables and is algorithm-independent (at least in the case of $e^+e^-\to 2$ jets). The function $S_{\mathrm{NG}}^{\mathrm{algo}}$ numerically resums the NGLs in the large-$N_c$ limit and depends on the algorithm. The function $ C_{\mathrm{clus}}^{\mathrm{algo}}$ resums CLs and is written as:
\begin{equation}\label{eq:YDclus}
 C_{\mathrm{clus}}^{\mathrm{algo}} = \exp\left[ \sum_{n=2}^{\infty} \frac{1}{n!} F_n^{\mathrm{algo}} (R) \left(-2C_Ft\right)^n \right],
\end{equation}
where $t=-\frac{1}{4\pi\beta_0} \ln(1-2\alpha_s\beta_0 L)$ and $\beta_0$ is the one-loop coefficient of the QCD $\beta$ function. The algorithm and radius-dependent coefficients $F_n^{\mathrm{algo}}$ result from the integration over rapidity and azimuth of the primary-emission $n$-loop amplitude over the relevant restricted phase-space.

\section{Results and discussion}

\begin{figure}
\includegraphics[width=.5\textwidth]{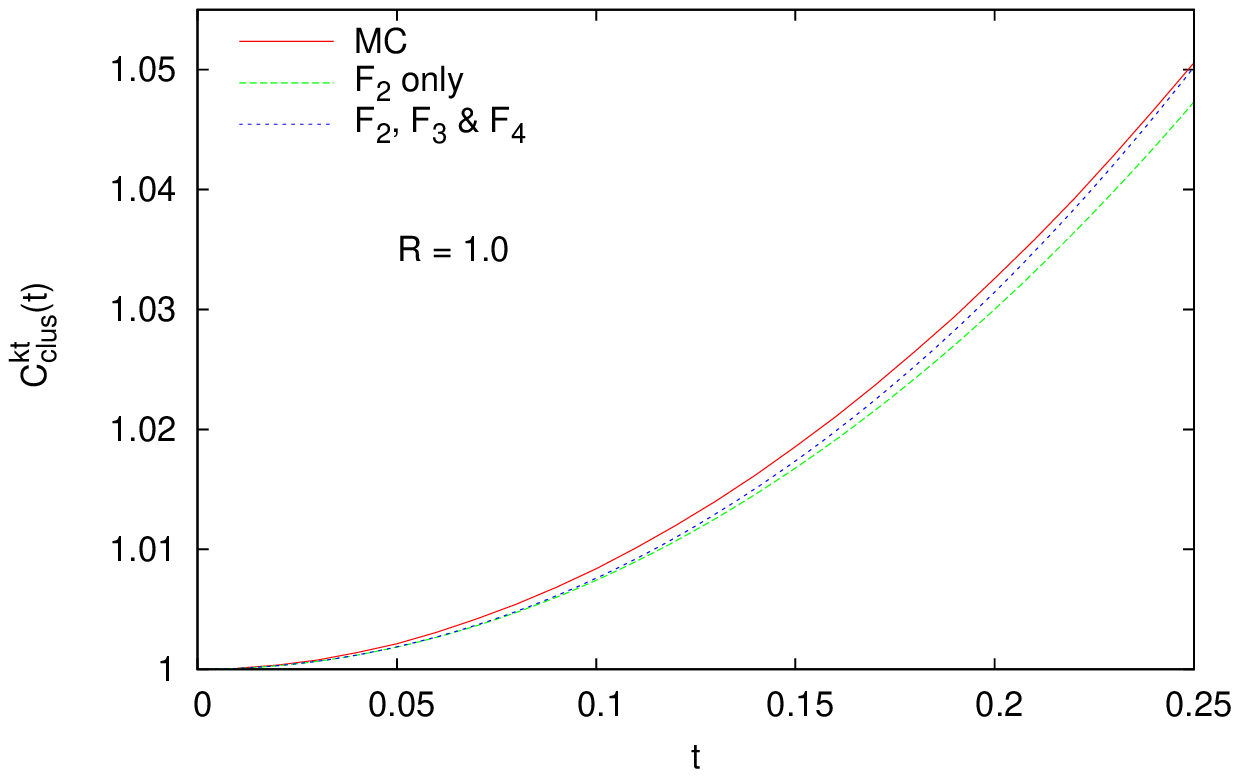}\hfill\includegraphics[width=.5\textwidth]{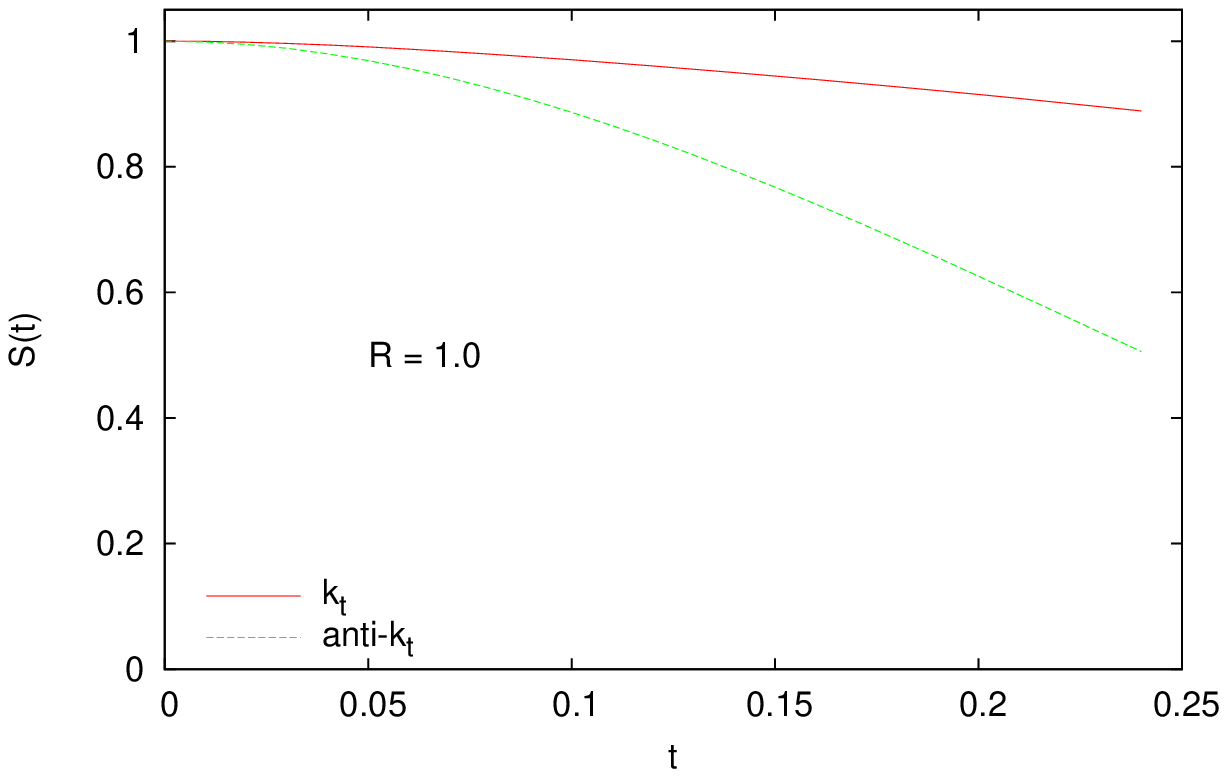}
\caption{\label{Yazid_Delenda_DIS2013_result} Plots of the functions $C_{\mathrm{clus}}^{k_t}$ (left) and $S_{\mathrm{NG}}$ (right) as indicated.}
\end{figure}

We show on the left-hand side of Fig. \ref{Yazid_Delenda_DIS2013_result} the function $C_{\mathrm{clus}}^{k_t}$, the resummed CLs form factor for the $k_t$ algorithm including up to four-loop clustering coefficients ($F_2^{k_t},\, F_3^{k_t},\, F_4^{k_t}$), compared to the output of the MC program of refs. \cite{ngl1, Appleby:2002ke, Delenda:2006nf}. We note the small impact of CLs form factor being of maximum order $5\%$ increase. This piece is totally absent in the anti-$k_t$ algorithm ($C_{\mathrm{clus}}^{\mathrm{anti}-k_t}=1$).
We also show on the right-hand side of the same figure the non-global function $S_{\mathrm{NG}}$ for both $k_t$ and anti-$k_t$ algorithms. As is clear from the plot the function $S_{\mathrm{NG}}$  is merely a factor $1\sim 0.9$ that multiplies the global and clustering form factors in the case of the $k_t$ algorithm, while it is a huge $\mathcal{O}(50\%)$ reduction factor in the case of the anti-$k_t$ algorithm.

The phenomenological implications of this result lead us to the concept of optimal jet algorithm and jet radius. While the anti-$k_t$ is often preferred over other algorithms in jet studies, it should be noticed that this is not always the best choice. One should instead be flexible in employing various jet algorithms and jet parameters (such as jet radius), as they may significantly affect the final results (Fig. \ref{Yazid_Delenda_DIS2013_result}). For example, given that the size of NGLs in the anti-$k_t$ algorithm is such a large reduction factor implies that neglected $\mathcal{O}(1/N_c)$ effects could (in principle) contribute up to order $5\%$ and maybe more, while their counterpart in the $k_t$ algorithm would contribute less than $1\%$, leading to more accurate predictions. The said accuracy is not, as might be expected,  spoiled by the presence of CLs because they are merely a factor of $5\%$ for which errors are not expected to contribute at more than the level of $0.5\%$. Furthermore, choosing larger jet radii means further reduction in NGLs while leaving CLs under control.  While this is true in the simple and clean $e^+e^-$ annihilation environment, the situation at hadron colliders is much more delicate. For instance, the UE in the latter environment scales as $R^2$ for many event shapes. In other words, larger jet radii result in greater UE contaminations and hence larger uncertainties.  Further discussion about this issue would be made more sensible once a resummation of CLs for hadronic collisions is performed, a task which is in progress.

\end{document}